\documentclass[bibyear]{aa}  

\usepackage{graphicx}
\usepackage{txfonts}
\begin{document}

   \title{An overlooked brown dwarf neighbour (T7.5 at $d$$\sim$5~pc) of the Sun and two additional T dwarfs at about 10~pc\thanks{based on observations with the Large Binocular Telescope (LBT)}}


\titlerunning{An overlooked brown dwarf neighbour}

   \author{G. Bihain
          \and
          R.-D. Scholz
          \and
          J. Storm
          \and
          O. Schnurr
          }

   \institute{Leibniz-Institut f\"ur Astrophysik Potsdam (AIP),
              An der Sternwarte 16, 14482 Potsdam, Germany\\
              \email{gbihain@aip.de, rdscholz@aip.de,
                     jstorm@aip.de, oschnurr@aip.de}
             }

   \date{Received 25 June 2013; accepted 10 July 2013 }

 
  \abstract
   {Although many new brown dwarf (BD) neighbours have recently been discovered
   thanks to new sky surveys in the mid- and near-infrared (MIR, NIR), their numbers are still more
   than five times lower than those of stars in the same volume.}
   {Our aim is to detect and classify new BDs to eventually
    complete their census in the immediate solar neighbourhood.}
   {We combined multi-epoch data from sky surveys at 
    different wavelengths to detect BD neighbours of the Sun by their high proper motion
    (HPM). We concentrated on relatively bright MIR ($w2$$<$13.5)
    BD candidates from the Wide-field Infrared Survey
    Explorer (WISE) expected to be so close to the Sun 
    that they may also be seen in older NIR (Two Micron All Sky Survey (2MASS)
    DEep Near-Infrared Survey (DENIS)) 
    or even red optical (Sloan Digital Sky Survey (SDSS) $i$- and $z$-band, 
    SuperCOSMOS Sky Surveys (SSS) $I$-band) surveys. 
    With low-resolution
    NIR spectroscopy we classified the new BDs and estimated their 
    distances and velocities.}
   {We have discovered the HPM ($\mu$$\sim$470~mas/yr) T7.5 dwarf, 
    WISE~J0521$+$1025, which is at $d$=5.0$\pm$1.3~pc from the Sun 
    the nearest known T dwarf in the northern sky, and two early T
    dwarfs, WISE~J0457$-$0207 (T2) and WISE~J2030$+$0749 (T1.5), 
    with proper motions of $\sim$120 and $\sim$670~mas/yr and 
    distances of 12.5$\pm$3.1~pc and 10.5$\pm$2.6~pc, respectively.
    The last one was independently discovered
    and also classified as a T1.5 dwarf by Mace and coworkers.
    All three show thin disc kinematics.
    They may have been overlooked in the past owing to overlapping
    images and because of problems with matching objects between different 
    surveys 
    and measuring their proper motions.}
   {}


   \maketitle


\section{Introduction}

The progress in discovering brown dwarfs (BDs) with ever cooler temperatures, that 
correspond to four spectral classes  (M, L, T, and Y),
is closely connected with the shift of all-sky surveys to longer wavelengths,
from the optical, to the near- and mid-infrared (NIR, MIR). 
As BDs change their spectral types
during their lifetime when cooling down (Burrows et al.~\cite{burrows01}), the
majority of BDs in the solar neighbourhood with typical ages of several Gyr are
expected to be T- and Y-type BDs. This has now been confirmed by the latest observations.

Updating the stellar and substellar census within 8~pc from the Sun
after the recently completed MIR WISE survey (Wide-field Infrared Survey 
Explorer; Wright et al.~\cite{wright10}), Kirkpatrick et al.~(\cite{kirkpatrick12})
listed 3 L-type, 22 T-type, and 8 Y-type objects. The last class was
only recently established by Cushing et al.~(\cite{cushing11}) and consists 
exclusively of WISE discoveries and will certainly be filled with many more 
discoveries. 
The WISE survey detected 7$+$1 new T and L dwarfs, respectively, in this volume, whereas
former NIR surveys, the Two Micron All Sky Survey
(2MASS; Skrutskie et al.~\cite{skrutskie06})
and the DEep Near-Infrared Survey (DENIS; Epchtein et al.~\cite{epchtein97}),
contributed 8$+$1 and 1$+$1 T and L dwarfs, respectively. Six T dwarfs
were found by other surveys, according to their discovery names listed in
Kirkpatrick et al.~(\cite{kirkpatrick12}). 

   \begin{figure*}
   \sidecaption
   \includegraphics[width=12.2cm]{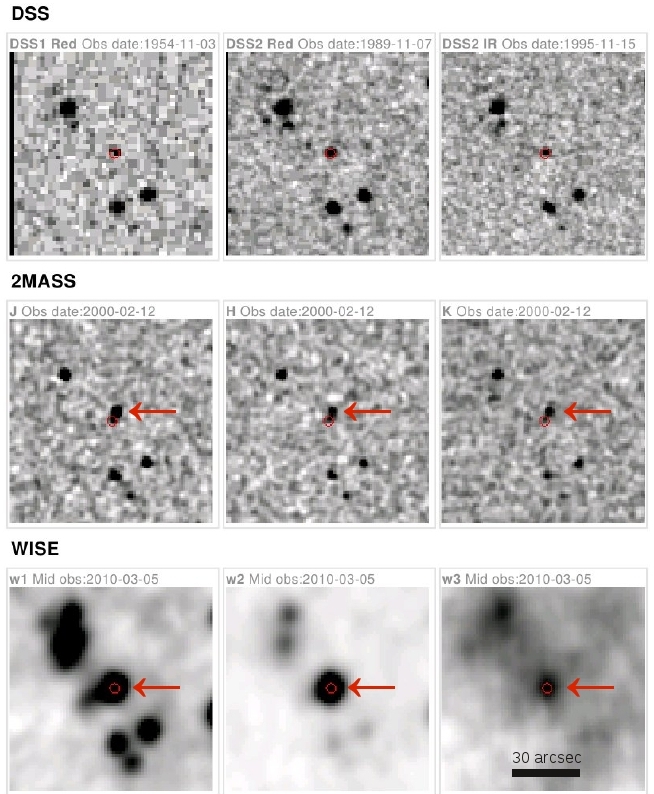}
      \caption{Digitised Sky Surveys (DSS), 2MASS, and WISE finding charts
               (90$\times$90~arcsec$^2$, north is up, east to the left)
               for WISE~J0521$+$1025. The red open circle marks the position
               of the target at the WISE epoch (i.e. the centre of each image). The
               red arrows mark the bright WISE source and the correct
               counterparts in 2MASS and DSS (if detected)
               of the HPM object. In this particular case, we see an
               overlapping background object in both 2MASS and DSS (within the
               open circle), whereas the correct counterpart is only seen in
               2MASS as the brighter object north of the background object.
               Compared to
               the other finding charts (Figs.~\ref{fig2_fc} and \ref{fig3_fc}),
               a higher magnification was chosen to better show the source confusion
               in this case.
              }
         \label{fig1_fc}
   \end{figure*}

Because of the small number density of L dwarfs and the optical faintness of T dwarfs,
none of the L/T discoveries from the Sloan Digital Sky Survey (SDSS) with its ongoing data 
releases (e.g. Abazajian et al.~\cite{abazajian09}, Aihara et al.~\cite{aihara11}) 
fall into the 8~pc sample, but one peculiar L6p/T7.5 binary, SDSS~J1416$+$1348AB 
(Bowler, Liu \& Dupuy \cite{bowler10}, Scholz~\cite{scholz10a}, 
Burgasser, Looper \&
Rayner~\cite{burgasser10a}), is missing according to the information given in
the DwarfArchives (Gelino, Kirkpatrick \& Burgasser~\cite{gelino12}). However,
the new accurate trigonometric parallax of this binary determined by 
Dupuy \& Liu~(\cite{dupuy12}) placed it at 9.11~pc, clearly outside the 8~pc horizon.
Only the nearest ($d$=3.626~pc) early T dwarf binary, $\varepsilon$ Indi Ba,Bb 
(Scholz et al.~\cite{scholz03}, McCaughrean et al.~\cite{mjm04}), was originally 
discovered in the optical as an unresolved high proper motion (HPM) object 
using two $I$-band photographic Schmidt plates with an epoch difference of several years
that were scanned within the SuperCOSMOS Sky Surveys 
(SSS; Hambly et al.~\cite{hambly01}). Also clearly seen on photographic Schmidt plates is 
the unresolved pair WISE~J1049$-$5319AB of two late-L dwarfs detected at the 
record-breaking distance of only 2~pc (Luhman~\cite{luhman13}, Mamajek~\cite{mamajek13}).

Kirkpatrick et al.~(\cite{kirkpatrick12}) 
found that there are currently about six times more stars than BDs within 8~pc.
They also expressed their expectation that this factor will decrease with time as
new discoveries are catalogued, and Luhman~(\cite{luhman13}) provided the first evidence
that these expectations are justified. His discovery was based on an HPM survey
taking advantage of the WISE data obtained in different seasons (with a mission lifetime 
of 13 months) and subsequent comparison with other surveys. Note that Luhman's
object was possibly overlooked in previous BD searches using 2MASS and DENIS,
and even photographic Schmidt plates, bacause of image crowding and resulting problems 
with the cross-matching of measured objects from different surveys.

Our BD search is also based on the identification of HPM objects; we 
first use WISE colour criteria and magnitude cuts and then check the candidates for 
shifted counterparts in other surveys with different epochs. This allowed us 
to detect two very nearby ($d$$\sim$5~pc) late T dwarfs (Scholz et al.~\cite{scholz11})
when the preliminary WISE data release first became available. Now we have used the
WISE All-Sky data release with similar selection criteria and have paid special 
attention to possible
mismatches with other surveys, which may prevent us from finding the correct counterparts.
Three newly found nearby BDs, one of which is a previously overlooked close
neighbour, are presented in this paper.


\section{Candidate selection and cross-identification}
\label{Cselpm}

We used the WISE All-Sky source catalogue with a mean observing epoch in
the first half of 2010 for the selection of bright MIR candidates with colours
typical of T dwarfs and hints on their possible HPM 
according to their cross-identification with 2MASS (epoch $\sim$2000) sources:

\begin{itemize}
\item Candidates were selected to have [$w1$$-$$w2$$>$1.5 (later than $\sim$T5)
and $w2$$<$13.5] or [0.5$<$$w1$$-$$w2$$<$1.5 ($\sim$T0-T5) and $w2$$<$12.5], aiming
at nearby ($d$$<$15~pc) T or Y dwarfs according to Figs.~1 and 29 in Kirkpatrick
et al.~(\cite{kirkpatrick11}).
\item To reduce crowding effects, only point sources outside the Galactic
plane ($|b|$$>$5$\degr$) were included.
\item To exclude extragalactic sources, only those with $w2$$-$$w3$$<$2.5
were considered (see Wright et al.~\cite{wright10}).
\item Only objects without a 2MASS counterpart (within 3~arcsec) or with a counterpart's
separation between 1~arcsec and 3~arcsec were selected as potential HPM candidates.
\end{itemize}

With the first two conditions we relied on the WISE MIR photometry of point sources,
which may however be affected by saturation for the brightest objects and by 
overlapping background objects not resolved by WISE, and effectively excluded most 
of the earlier-type BDs and stars from our target list. As we applied a relatively 
bright WISE magnitude cut, we expected to see these objects also in the 2MASS,
if they were not as cool as Y dwarfs. Therefore, our fourth condition was aimed at finding
either HPM objects with $\mu$$>$0.3~arcsec/yr or with 0.1$<$$\mu$$<$0.3~arcsec/yr
given the WISE-2MASS epoch difference of about ten years. However, we considered
the 2MASS counterparts with 1-3~arcsec shifts as suspicious and wanted to 
visually inspect
the corresponding WISE sources for alternative HPM counterparts
outside of the search radius of 3~arcsec.

\begin{figure}
   \centering
   \includegraphics[width=9.2cm]{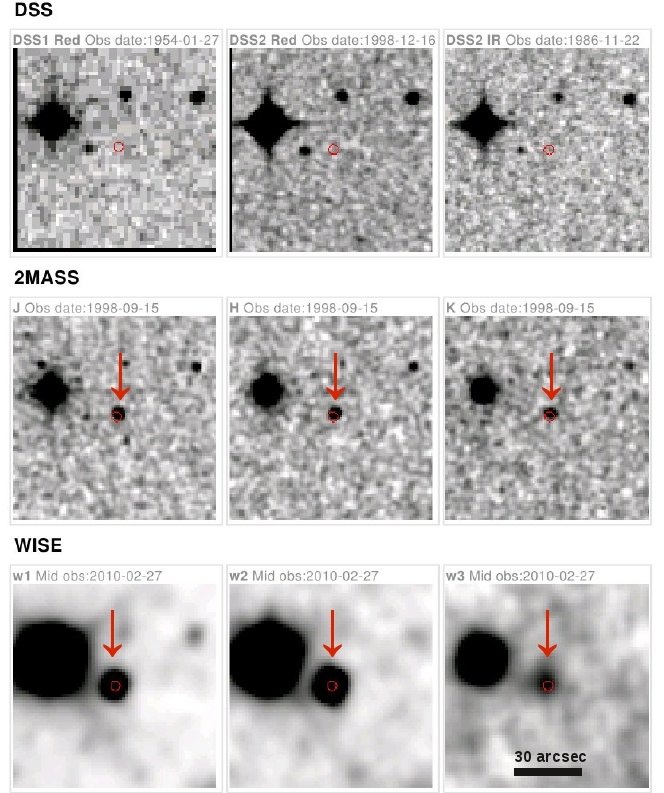}
      \caption{Finding charts as in Fig.~\ref{fig1_fc}
               for WISE~J0457$-$0207.
              }
         \label{fig2_fc}
   \end{figure}

\begin{figure}
   \centering
   \includegraphics[width=9.2cm]{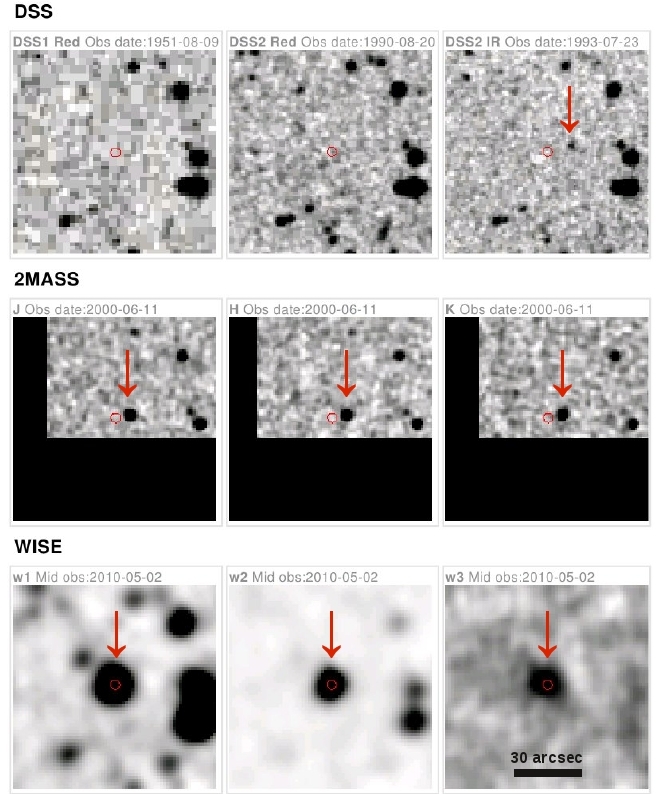}
      \caption{Finding charts as in Fig.~\ref{fig1_fc} 
	       for WISE~J2030$+$0749.
              }
         \label{fig3_fc}
   \end{figure}

About 2000 candidates were found with the above conditions. With the help of the
IRSA Finder Charts tool\footnote{http://irsa.ipac.caltech.edu/applications/finderchart/
providing DSS, 2MASS, and WISE 
images for a given object at a glance (see e.g. Figs.~\ref{fig1_fc}-\ref{fig3_fc}}),
we were able to inspect all these candidates to identify HPM objects.
These were then checked for known objects in 
DwarfArchives (Gelino, Kirkpatrick \& Burg\-asser~\cite{gelino12}) and 
SIMBAD\footnote{http://simbad.u-strasbg.fr/}.
Although most of the 2000 initial candidates
were rejected as ghosts/stripes, reddened or extended/diffuse objects, we
found some variable stars (e.g. a new Galactic Nova; 
Scholz et al.~\cite{scholz12b}) 
and many previously known BD and stellar neighbours of the Sun: more than 40
T dwarfs, about 20 L dwarfs, but also about 20 M dwarfs and earlier-type stars.
Among about ten new candidates, we selected three with photometrically estimated 
distances of less than about 10~pc and moderately low declinations
for spectroscopic follow up (see Sect.~\ref{NIRsp}) with the 
Large Binocular Telescope (LBT) (other early T-type 
and red L-type candidates
were placed in different observing programmes and will be published elsewhere).
We matched them
with 2MASS and also with later WISE observations, and two could be
identified in other NIR/optical surveys as well
(Table~\ref{table:1}). Finally,
we used the recently measured positions of our targets on the LBT acquisition
images (Sect.~\ref{NIRsp}) calibrated with the PPMXL (R\"oser, Demleitner \& Schilbach~\cite{roeser10})
to confirm the proper motions and improve their accuracy.

\textbf{WISE~J052126.29$+$102528.4} \\
(hereafter WISE~J0521$+$1025) - For this late T candidate ($w1$$-$$w2$=$+$1.8)
the WISE catalogue lists a 2MASS counterpart 
separated by 1.4~arcsec. This is obviously a background object 
that is also visible in the DSS (Fig.~\ref{fig1_fc}). However,
the brighter 2MASS object north of it appears blue in the NIR
and has no optical counterpart, indicating already on the basis of the
2MASS data alone a HPM T-type BD candidate.
Both objects are flagged in the 2MASS as deblended in $J$ and $K_s$, and as
the astrometry may also be affected, we measured the 2MASS position of the 
blue object visually using the ESO Skycat tool. We
also found a second epoch in the WISE 3-band cryo data 
(Table~\ref{table:1}).

\textbf{WISE~J045746.08$-$020719.2} \\
(hereafter WISE~J0457$-$0207) - In this case, the 2MASS counterpart shifted
by 1.6~arcsec is not seen in the optical (Fig.~\ref{fig2_fc})
and is moderately red, $J$$-$$K_s$=$+$0.9, consistent with an 
early T dwarf with a relatively small proper motion. The colours $w1$$-$$w2$=$+$1.0
and $J$$-$$w2$=$+$2.5 agree with this classification. In addition, this 
object is detected by DENIS and by the Galactic Clusters Survey (GCS) within 
the UKIRT InfraRed Deep Sky Surveys
(UKIDSS)\footnote{The UKIDSS project is defined in 
Lawrence et al.~(\cite{lawrence07}).
UKIDSS uses the UKIRT Wide Field Camera (WFCAM; Casali et al.~\cite{casali07})
and the photometric system described in Hewett et al.~(\cite{hewett06}),
which is situated in the Mauna Kea Observatories (MKO) system (Tokunaga et
al.~\cite{tokunaga02}).
The pipeline processing and science archive are described
in Hambly et al.~(\cite{hambly08}) and Irwin et al.~(in prep.).}. Later we
found another detection in the WISE 3-band cryo data 
(Table~\ref{table:1}).

\textbf{WISE~J203042.79$+$074934.7} \\
(hereafter WISE~J2030$+$0749) - No 2MASS counterpart ($<$3~arcsec) was listed 
for this one, but the finding charts in Fig.~\ref{fig3_fc} show a clear
HPM object with growing separation from 2MASS to older DSS IR. From the
SSS we found three $I$-band positions, and the object was
also detected in the SDSS $iz$ bands (Table~\ref{table:1}). 
Its colours ($i$$-$$z$=$+$4.6, 
$J$$-$$K_s$=$+$0.9, $J$$-$$w2$=$+$2.1, $w1$$-$$w2$=$+$0.8) 
fit a T2 dwarf (Hawley et al.~\cite{hawley02},
Kirkpatrick et al.~\cite{kirkpatrick11}). However, there is only 
one T2 dwarf listed in Hawley et al.~(\cite{hawley02}) that has $i$$-$$z$=$+$4.2,
whereas the average values of $<$T2 and $>$T2 dwarfs are generally smaller and reach
$i$$-$$z$=$+$4.0 only for the latest-given class of T6 dwarfs.
From WISE post-cryo single
exposures we determined an additional mean position at a
later epoch (Table~\ref{table:1}).

%
\begin{table}
\caption{Positions (J2000), proper motions [mas/yr], photometry [mag], spectral indices/types, distances [pc], and  tangential velocities [km/s]} 
\label{table:1}      
\centering                          
\begin{tabular}{lccc}        
\hline\hline                 
Param.   & J0521$+$1025                            & J0457$-$0207      & J2030$+$0749   \\ 
\hline                        
\tiny{LBT $\alpha$} &                     05 21 26.349       &   04 57 46.114    &   20 30 42.897  \\ 
\tiny{LBT $\delta$} &                   $+$10 25 27.41       & $-$02 07 19.59    & $+$07 49 34.44  \\
\tiny{LBT} ep       &                          2012.773      &        2012.770   &        2012.855 \\
\tiny{WISE} $\alpha$ &                    05 21 26.2967      &   04 57 46.0884   &   20 30 42.7986 \\
\tiny{WISE} $\delta$ &                  $+$10 25 28.494      & $-$02 07 19.239   & $+$07 49 34.741 \\ 
\tiny{WISE} ep       &                         2010.175      &        2010.156   &        2010.332 \\
\tiny{WISE}\tablefootmark{c} $\alpha$ &   05 21 26.3165      &   04 57 46.1024   &   20 30 42.8069 \\
\tiny{WISE}\tablefootmark{c} $\delta$ & $+$10 25 28.439      & $-$02 07 19.186   & $+$07 49 34.602 \\  
\tiny{WISE}\tablefootmark{c} ep   &            2010.701      &        2010.682   &        2010.830 \\
\tiny{GCS}\tablefootmark{b} $\alpha$ & n/a                   &   04 57 46.0785   & n/a \\
\tiny{GCS}\tablefootmark{b} $\delta$ & n/a                   & $-$02 07 19.202   & n/a \\
\tiny{GCS} ep       &                  n/a                   &        2010.019   & n/a \\
\tiny{SDSS} $\alpha$ &                 n/a                   &  n/a              &    20 30 42.357 \\
\tiny{SDSS} $\delta$ &                 n/a                   &  n/a              &  $+$07 49 35.64 \\
\tiny{SDSS} ep       &                 n/a                   &  n/a              &        2000.748 \\
\tiny{2MASS} $\alpha$ &        05 21 26.147\tablefootmark{a} &    04 57 46.022   &    20 30 42.357 \\
\tiny{2MASS} $\delta$ &      $+$10 25 32.74\tablefootmark{a} &  $-$02 07 17.95   &  $+$07 49 35.83 \\
\tiny{2MASS} ep       &                        2000.118      &        1998.707   &        2000.444 \\
\tiny{DENIS} $\alpha$ &         n/a                          &    04 57 46.038   &  n/a  \\
\tiny{DENIS} $\delta$ &         n/a                          &  $-$02 07 18.34   &  n/a  \\
\tiny{DENIS} ep       &         n/a                          &        1998.953   &  n/a  \\
\tiny{SSS} $I$                  $\alpha$ & n/d               &  n/d              &    20 30 42.149 \\
\tiny{SSS} $I$                  $\delta$ & n/d               &  n/d              &  $+$07 49 36.18 \\
\tiny{SSS} $I$                  ep    &    1995.874          &  2001.862         &        1995.654 \\
\tiny{SSS} $I$\tablefootmark{f} $\alpha$ & n/a               &  n/a              &    20 30 42.051 \\
\tiny{SSS} $I$\tablefootmark{f} $\delta$ & n/a               &  n/a              &  $+$07 49 37.35 \\
\tiny{SSS} $I$\tablefootmark{f} ep    &    n/a               &  n/a              &        1993.545 \\
\hline
$\mu_{\alpha}\cos{\delta}$ &           $+$232$\pm$9  &       $+$82$\pm$9 &    $+$653$\pm$6 \\
$\mu_{\delta}$             &           $-$418$\pm$6  &       $-$97$\pm$8 &    $-$138$\pm$16 \\
\hline
\tiny{SSS} $I$     &                 n/d                     &  n/d              &   $\sim$19.5 \\
\tiny{SSS} $I$\tablefootmark{f}     &  n/a                   &  n/a              &   $\sim$18.9$\pm$0.3 \\
\tiny{SDSS} $i$    &                 n/a                     &  n/a              &  21.810$\pm$0.140 \\
\tiny{SDSS} $z$    &                 n/a                     &  n/a              &  17.195$\pm$0.014 \\
\tiny{DENIS} $J$   &                 n/a                     &  14.879$\pm$0.12  &  n/a \\
\tiny{2MASS} $J$   &                 15.262\tablefootmark{d} &  14.897$\pm$0.040 &  14.227$\pm$0.029 \\
\tiny{2MASS} $H$   &                        15.222$\pm$0.103 &  14.198$\pm$0.046 &  13.435$\pm$0.033 \\
\tiny{2MASS} $K_s$ &                 14.665\tablefootmark{d} &  14.022$\pm$0.055 &  13.319$\pm$0.039 \\
\tiny{GCS} $H$\tablefootmark{e}     &       n/a              &  14.190$\pm$0.003 &  n/a \\
\tiny{GCS} $K$\tablefootmark{e}     &       n/a              &  13.975$\pm$0.003 &  n/a \\
\tiny{WISE} $w1$   &                        14.098$\pm$0.031 &  13.391$\pm$0.026 &  12.956$\pm$0.025 \\
\tiny{WISE} $w2$   &                        12.286$\pm$0.026 &  12.443$\pm$0.025 &  12.122$\pm$0.025 \\                      
\tiny{WISE} $w3$   &                        10.306$\pm$0.085 &  11.020$\pm$0.114 &  10.964$\pm$0.110 \\
\hline
\tiny{H$_2$O-$H$}  & 0.246 (T7)    & 0.509 (T2)    & 0.599 (T0/T1)\\
\tiny{CH$_4$-$H$}  & 0.155 (T7/T8) & 0.798 (T2/T3) & 0.859 (T2)   \\
\tiny{CH$_4$-$K$}  & 0.084 (T7/T8) & 0.488 (T3)    & 0.595 (T2)   \\
\tiny{SpT$_{vis}$}   & T7.5        & T2            & T1.5         \\
\tiny{SpT$_{adopt}$} & T7.5        & T2            & T1.5         \\
$d$         & 5.0$\pm$1.3   & 12.5$\pm$3.1  & 10.5$\pm$2.6 \\
$v_{tan}$   & 11$\pm$3      & 8$\pm$2       & 33$\pm$8     \\
\hline                                   
\end{tabular}
\tablefoot{
Basic WISE data are from the All-Sky source catalogue, 
SDSS data from DR8 (Aihara et al.~\cite{aihara11}). 
For other data references, see text (n/a - not available, n/d - not detected).
\tablefoottext{a}{visual measurement in 2MASS FITS image}
\tablefoottext{b}{mean coordinates from four measurements at one epoch}
\tablefoottext{c}{WISE 3-band cryo source working database or mean of Post-Cryo Single Exposure (L1b) data (for J2030$+$0749)}
\tablefoottext{d}{2MASS image deblended in this band}
\tablefoottext{e}{mean GCS aperMag3 magnitudes from two measurements}
\tablefoottext{f}{mean from two additional epochs - 1993.533$+$1993.557}
}
\end{table}

   \begin{figure*}
   \centering
   \includegraphics[width=14.0cm]{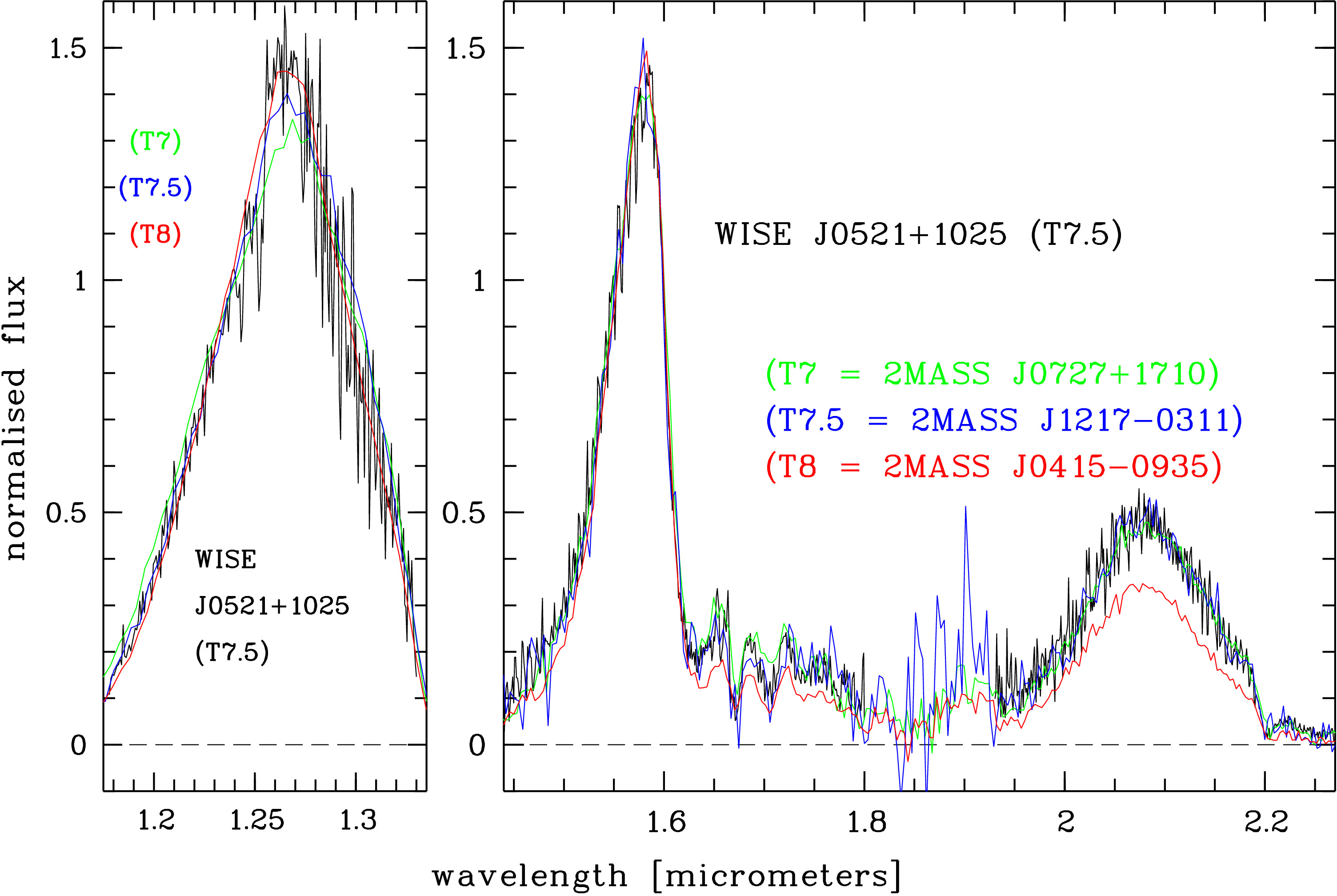}
      \caption{LBT/LUCI $J$-band (left) and $HK$-band (right) spectra
of WISE~J0521$+$1025 (black)
overplotted with lower resolution standard spectra of
2MASS~J0727$+$1710 (T7, green),
2MASS~J1217$-$0311 (T7.5, blue) (Burgasser et al.~\cite{burgasser06}), and
2MASS~J0415$-$0935 (T8, red) (Burgasser et al.~\cite{burgasser04}).
              }
         \label{fig4_spec}
   \end{figure*}

   \begin{figure*}
   \centering
   \includegraphics[width=14.0cm]{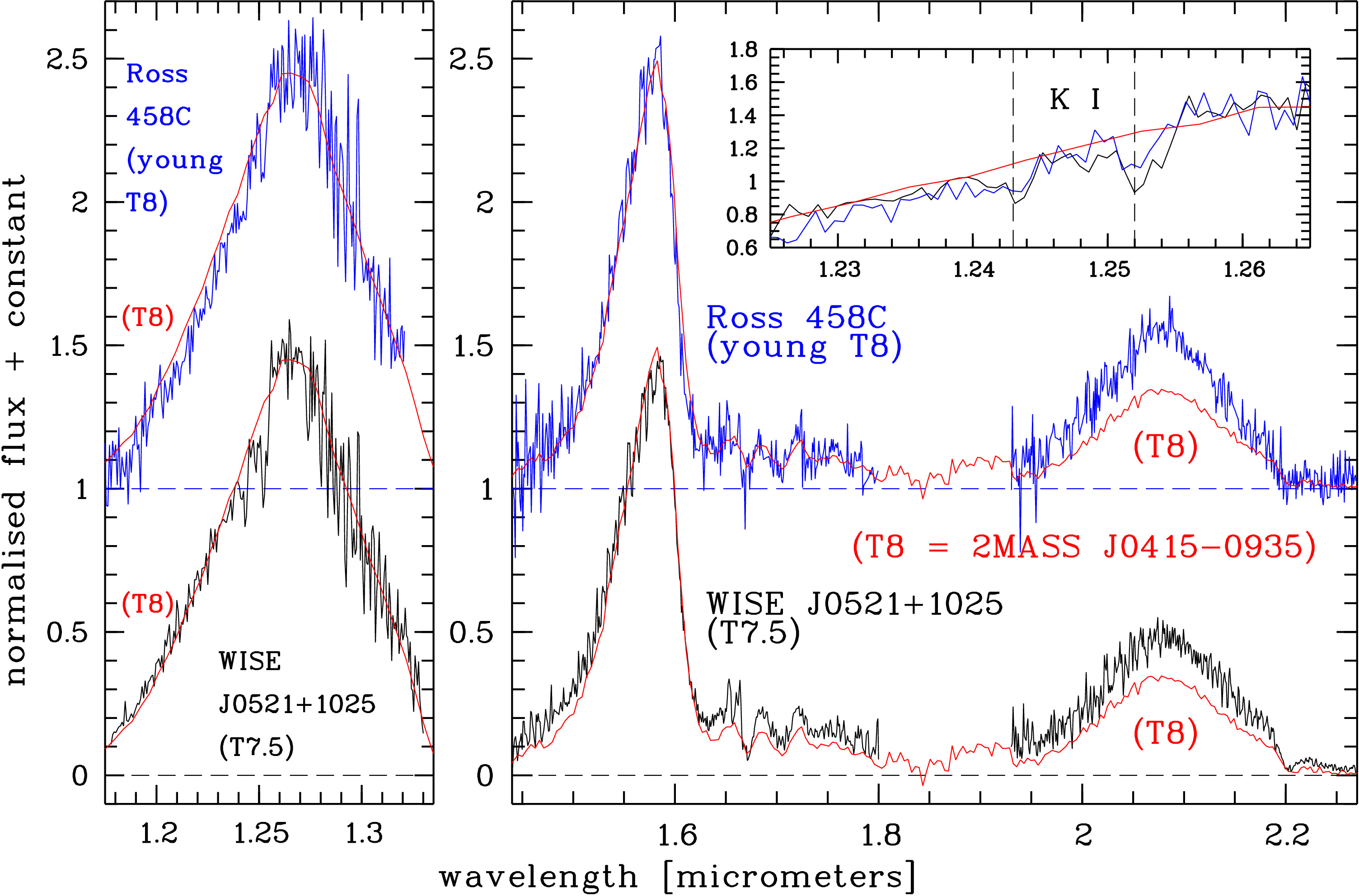}
      \caption{LBT/LUCI spectra
of WISE~J0521$+$1025 (black) and Ross~458C (blue, from Scholz et al.~\cite{scholz11})
overplotted with lower resolution spectrum (red) of the T8 standard
2MASS~J0415$-$0935 (Burgasser et al.~\cite{burgasser04}). The insert
shows the region of the K~I doublet at 1.243/1.252~$\mu$m.
              }
         \label{fig5_spec}
   \end{figure*}

   \begin{figure*}
   \centering
   \includegraphics[width=14.0cm]{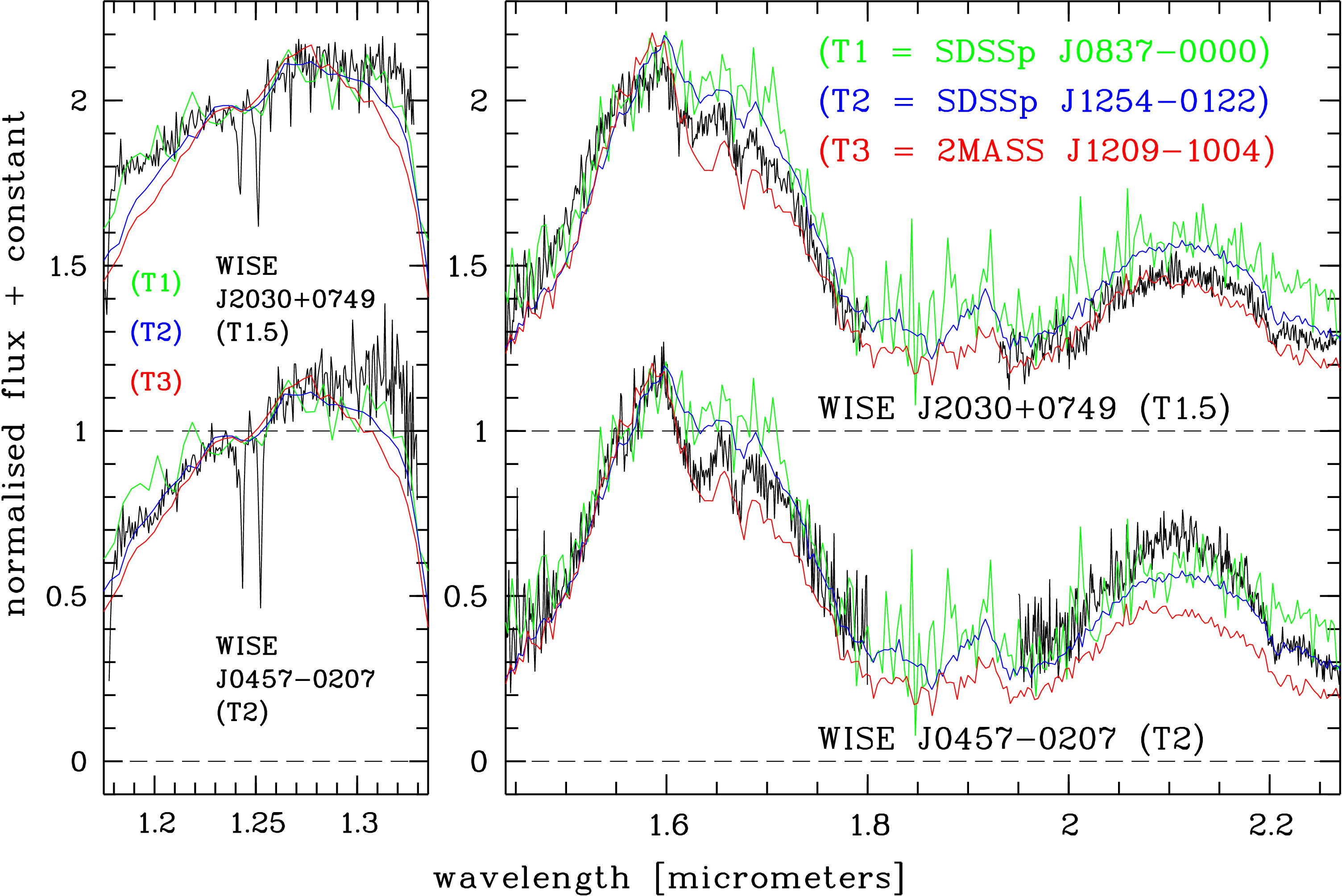}
      \caption{LBT/LUCI spectra (black)
of WISE~J0457$-$0207 and WISE~J2030$+$0749
overplotted with lower resolution standard spectra:
SDSSp~J0837$-$0000 (T1, green) (Burgasser et al.~\cite{burgasser06}),
SDSSp~J1254$-$0122 (T2, blue), and
2MASS~J1209$-$1004 (T3, red)
(Burgasser et al.~\cite{burgasser04}).
              }
         \label{fig6_spec}
   \end{figure*}


\section{Near-infrared spectroscopic classification}
\label{NIRsp}

Our three targets were observed with the
LBT NIR spectrograph LUCI~1
(Mandel et al.~\cite{mandel08}; Seifert et al.~\cite{seifert10};
Ageorges et al.~\cite{ageorges10}) in long-slit
spectroscopic mode with
the $HK$ (200 lines/mm + order separation filter)
and $zJHK$ gratings (210 lines/mm + $J$ filter).
The dwarf WISE~J0521$+$1025 was observed on 2012-Oct-09 with total integration
times of 40~min in $HK$ and 20~min in $J$,  WISE~J0457$-$0207 and
WISE~J2030$+$0749 on 2012-Oct-08 and 2012-Nov-08, respectively,
but both with only 16~min ($HK$) and 10~min ($J$). As in
Scholz et al.~(\cite{scholz11,scholz12a}), central
wavelengths were chosen at 1.835~$\mu$m ($HK$) and
1.25~$\mu$m ($J$) yielding a coverage of 1.38--2.26 and
1.18--1.33~$\mu$m, respectively. The slit width was always
1~arcsec, corresponding to a spectral resolving power of
$R$=$\lambda$/$\Delta$$\lambda$$\approx$4230, 940,
and 1290 at $\lambda$$\approx$1.24,
1.65, and 2.2~$\mu$m, respectively.
Observations consisted of individual
exposures of 60~s in $HK$ (75~s for WISE~J0521$+$1025) and 150~s in $J$
with shifting the target along the slit
using an ABBA pattern until the total integration time was reached.
For more details and a description of the spectroscopic data reduction
we refer the reader to Scholz et al.~(\cite{scholz11,scholz12a}).
Note that the above given wavelength coverage is not wide enough at both
the blue and red ends to compute spectrophotometric colours in the 2MASS
system (using spectral response curves from Cohen et al.~\cite{cohen03}).
The $J$ band is also too narrow to compute spectral indices
for classifying T dwarfs according to Burgasser et al.~(\cite{burgasser06})
so that only $HK$ indices can be used.

In Figs.~\ref{fig4_spec}, \ref{fig5_spec}, and \ref{fig6_spec},
we show $J$- and $HK$-band spectra normalised at 
1.2-1.3~$\mu$m and 1.52-1.61~$\mu$m, respectively. 
The $J$-band spectrum of WISE~J0521$+$1025 fits best to that of a
T8 standard, but is more similar to T7/T7.5 in the $HK$ band,
with a better fit to T7.5 at 1.7~$\mu$m (Fig.~\ref{fig4_spec}). 
Except for the $H$ band,
we note a good agreement, including the K~I doublet 
(at 1.24/1.25~$\mu$m) in the $J$ band
and the high peak in the $K$ band, with 
Ross 458C (discovered by Goldman et al.~\cite{goldman10} and Scholz~\cite{scholz10b}) 
observed with the same instrument (Fig.~\ref{fig5_spec}). 
Because of these features,
Ross 458C was characterised as a young (low surface gravity) 
and super-solar metallicity T8 dwarf
by Burgasser et al.~(\cite{burgasser10b}), 
whereas Burningham et al.~(\cite{burningham11}) typed it as T8.5p.
We visually classified WISE~J0521$+$1025 as T7.5 in good agreement with the measured
spectral indices in the $HK$ band (Table~\ref{table:1}) as defined 
in Burgasser et al.~(\cite{burgasser06}). 

The spectra of WISE~J0457$-$0207 (with a remarkably high $K$-band peak 
that cannot be explained by uncertainties of the flux calibration)
and WISE~J2030$+$0749 are 
of earlier ($\sim$T2) type (Fig.~\ref{fig6_spec}), 
fitting in parts better to the T1, T2,
or T3 standard. As standards are single,
this may indicate possible close binary components
with different types or peculiarities related to age or metallicity.
The extreme $i$$-$$z$ index of WISE~J2030$+$0749 makes this object even more interesting.
Visually we classified WISE~J0457$-$0207 as T2 and WISE~J2030$+$0749 as T1.5
and adopted these types consistent with those obtained
from spectral indices.

Using mean absolute WISE magnitudes of single T7.5 and T1/T2 dwarfs from 
Dupuy \& Liu~(\cite{dupuy12}), we estimated distances of 5.0$\pm$1.3~pc for WISE~J0521$+$1025, 
12.5$\pm$3.1~pc for WISE~J0457$-$0207, and 10.5$\pm$2.6~pc for WISE~J2030$+$0749.


\section{Conclusions}
\label{Sconc}

We have discovered three new BDs close to the Sun in an HPM
search using MIR, NIR, and optical surveys: WISE~J0457$-$0207
has a relatively small proper motion for an object at the 10~pc horizon (cf. Fig.~1 in
Scholz et al.~\cite{scholz11}) not detectable in the past because of
similar 2MASS and DENIS epochs. WISE~J2030$+$0749, with similar 2MASS and SDSS
epochs, was previously not associated with its SSS measurement, whereas
WISE~J0521$+$1025 was probably overlooked in previous BD and HPM
searches because of problems matching partly blended
images in different surveys.
      
Using NIR spectroscopy with LBT/LUCI we classified WISE~J0521$+$1025
as a new T7.5 dwarf at a distance of about 5~pc. It is currently the
nearest T dwarf in the northern hemisphere and 
may also be the closest free-floating neighbour of its spectral sub-class.
The dwarfs WISE~J0457$-$0207 and WISE~J2030$+$0749 lie,
according to their T2 and T1.5 types, slightly beyond 10~pc, but
may still fall in the 10~pc sample given their error bars, if they
are not unresolved binaries.
The latter was independently discovered by 
Mace et al.~(\cite{mace13}), who also classified it as a T1.5 dwarf.
However, they did not mention its large proper motion, 
proximity, and very red $i$$-$$z$ colour
from the SDSS.
The small tangential velocities of all three new BDs are typical of the 
Galactic thin disc population. They are promising targets for
trigonometric parallax programmes and adaptive optics observations.


\begin{acknowledgements}
The authors thank Jochen Heidt, Barry Rothberg,
and all observers at the LBT for assistance during the
preparation and execution of LUCI observations, 
Adam Burgasser for providing template spectra
at http://pono.ucsd.edu/$\sim$adam/browndwarfs/spexprism,
and the anonymous referee for a quick and helpful report
and Victor J. Sanchez Bejar for some important hints.

This research has made use of the WFCAM Science Archive
providing UKIDSS, the
NASA/IPAC Infrared Science Archive, which is operated by the Jet Propulsion
Laboratory, California Institute of Technology, under contract with the
National Aeronautics and Space Administration,
and of data products from WISE,
which is a joint project of the University of California,
Los Angeles, and the Jet Propulsion Laboratory/California Institute of
Technology, funded by the National Aeronautics and Space Administration,
from 2MASS, and from SDSS DR7 and DR8.
Funding for SDSS-III has been provided by the Alfred P. Sloan Foundation,
the Participating Institutions, the National Science Foundation, and the
U.S. Department of Energy. The SDSS-III web site is http://www.sdss3.org/.
This research has benefitted from the M, L, T, and Y dwarf compendium housed at 
DwarfArchives.org.
We have also used SIMBAD and VizieR at the CDS/Strasbourg.
\end{acknowledgements}



\begin{thebibliography}{}

\bibitem[2009]{abazajian09}
Abazajian, K.~N., Adelman-McCarthy, J.~K., Ag\"ueros, M.~A., et al.\ 2009, ApJS, 182, 54

\bibitem[2010]{ageorges10}
Ageorges, N., Seifert, W., J\"utte, M., et al.\ 2010, SPIE, 7735, 53

\bibitem[2011]{aihara11}
Aihara, H., Allende Prieto, C., An, D., et al.\ 2011, ApJS, 193, 29

\bibitem[2010]{bowler10}
Bowler, B.~P., Liu, M.~C., \& Dupuy, T.~J.\ 2010, ApJ, 710, 45

\bibitem[2001]{burrows01}
Burrows, A., Hubbard, W.~B., Lunine, J.~I., \& Liebert, J.\ 2001,
Reviews of Modern Physics, 73, 719

\bibitem[2006]{burgasser06}
Burgasser, A.~J., Geballe, T.~R., Leggett, S.~K., Kirkpatrick, J.~D.,
\& Golimowski, D.~A.\ 2006, \apj, 637, 1067

\bibitem[2010a]{burgasser10a}
Burgasser, A.~J., Looper, D., Rayner, J.~T.\ 2010a, AJ, 139, 2448

\bibitem[2010b]{burgasser10b}
Burgasser, A.~J., Simcoe, R.~A., Bochanski, J.~J., et al.\ 2010b, ApJ, 725, 1405

\bibitem[2004]{burgasser04}
Burgasser, A.~J., McElwain, M.~W., Kirkpatrick, J.~D., et al.\ 2004, AJ, 127, 2856

\bibitem[2011]{burningham11}
Burningham, B., Leggett, S.~K., Homeier, D., et al.\ 2011, MNRAS, 414, 3590

\bibitem[2007]{casali07}
Casali, M., Adamson, A., Alves de Oliveira, C., et al.\ 2007, A\&A, 467, 777

\bibitem[2003]{cohen03}
Cohen, M., Wheaton, W.~A., Megeath, S.~T. \ 2003, AJ, 126. 1090

\bibitem[2011]{cushing11}
Cushing, M.~C., Kirkpatrick, J.~D., Gelino, C.~R., et al.\ 2011, ApJ, 743, 50 

\bibitem[2012]{dupuy12}
Dupuy, T.~J., \& Liu, M.~C.\ 2012, ApJ Suppl. Ser., 201, 19

\bibitem[1997]{epchtein97}
Epchtein, N., de Batz, B., Capoani, L., et al.\ 1997, The Messenger, 87, 27

\bibitem[2012]{gelino12}
Gelino, C.~R., Kirkpatrick, J.~D., \& Burgasser, A.~J.\ 2012,
online database for 804 L and T dwarfs at DwarfArchives.org
(status: 6 November 2012)

\bibitem[2010]{goldman10}
Goldman, B., Marsat, S., Henning, T., Clemens, C., \& Greiner, J.\ 2010, MNRAS, 405, 1140

\bibitem[2001]{hambly01}
Hambly, N.~C.,  MacGillivray, H.~T., Read M.~A., et al.\ 2001, MNRAS, 326, 1279

\bibitem[2008]{hambly08}
Hambly, N.~C., Collins, R.~S., Cross, N.~J.~G., et al.\ 2008, MNRAS, 384, 637

\bibitem[2002]{hawley02}
Hawley, S. L., Covey, K. R., Knapp, G. R., et al.\ 2002, AJ, 123, 3409

\bibitem[2006]{hewett06}
Hewett, P.~C., Warren, S.~J., Leggett, S.~K., \& Hodgkin, S.~T.\ 2006,
MNRAS, 367, 454

\bibitem[2011]{kirkpatrick11}
Kirkpatrick, J.~D., Cushing, M.~C., Gelino, C.~R., et al.\ 2011, ApJ Suppl. Ser., 197, 19 

\bibitem[2012]{kirkpatrick12}
Kirkpatrick, J.~D., Gelino, C.~R., Cushing, M.~C., et al.\ 2012, ApJ, 753, 156

\bibitem[2007]{lawrence07}
Lawrence, A., Warren, S.~J., Almaini, O., et al.\ 2007, MNRAS, 379, 1599

\bibitem[2013]{luhman13}
Luhman, K.~L.\ 2013, ApJ, 767, L1

\bibitem[2013]{mace13}
Mace, G.~N., Kirkpatrick, J.~D., Cushing, M.~C., et al.\ 2013, 
ApJ Suppl. Ser, 205, 6

\bibitem[2013]{mamajek13}
Mamajek, E.~E.\ 2013, arXiv:1303.5345

\bibitem[2008]{mandel08}
Mandel, H., Seifert, W., Hofmann, R., et al.\ 2008, SPIE, 7014, 124

\bibitem[2004]{mjm04}
McCaughrean, M.~J., Close, L.~M., Scholz, R.-D., et al.\ 2004, A\&A, 413, 1029

\bibitem[2010]{roeser10}
R\"oser, S., Demleitner, M., Schilbach, E.\, 2010, AJ, 139, 2440

\bibitem[2003]{scholz03}
Scholz, R.-D., McCaughrean, M.~J., Lodieu, N., \& Kuhlbrodt, B.\ 2003, A\&A, 398, L29

\bibitem[2010a]{scholz10a}
Scholz, R.-D.\ 2010a, A\&A, 510, L8

\bibitem[2010b]{scholz10b}
Scholz, R.-D.\ 2010b, A\&A, 515, A92

\bibitem[2011]{scholz11}
Scholz, R.-D., Bihain, G., Schnurr, O., \& Storm, J.\ 2011, A\&A, 532, L5 

\bibitem[2012a]{scholz12a}
Scholz, R.-D., Bihain, G., Schnurr, O., \& Storm, J.\ 2012a, A\&A, 541, A163 

\bibitem[2012b]{scholz12b}
Scholz, R.-D., Granzer, 
T., Schwarz, R., et al.\ 2012b, The Astronomer's Telegram, 4268, 1

\bibitem[2010]{seifert10}
Seifert, W., Ageorges, N., Lehmitz, M., et al.\ 2010, SPIE, 7735, 256

\bibitem[2006]{skrutskie06}
Skrutskie, M.~F., Cutri, R.~M., Stiening, R., et al.\ 2006, AJ, 131, 1163

\bibitem[2002]{tokunaga02}
Tokunaga, A.~T., Simons, D.~A., \& Vacca, W.~D.\ 2002, PASP, 114, 180

\bibitem[2010]{wright10}
Wright, E.~L., Eisenhardt, P.~R.~M., Mainzer, A.~K., et al.\ 2010, AJ, 140, 1868

\end{thebibliography}
\end{document}